\documentclass[preprint]{elsarticle}

\usepackage{graphicx}
\usepackage{amssymb}
\usepackage{tabularx}
\usepackage{xcolor}
\journal{Journal of Magnetism and Magnetic Materials}

\begin{document}

\begin{frontmatter}

\title{The spontaneous exchange bias effect in La$_{2-x}$Ca$_{x}$CoMnO$_{6}$ series}

\author[AA]{J. R. Jesus}, \author[BB]{L. Bufai\c{c}al\corref{Bufaical}}, \ead{lbufaical@ufg.br}, \author[AA]{E. M. Bittar}

\address[AA]{Centro Brasileiro de Pesquisas F\'{\i}sicas, Rua Dr. Xavier Sigaud 150, 22290-180, Rio de Janeiro, RJ, Brazil}
\address[BB]{Instituto de F\'{i}sica, Universidade Federal de Goi\'{a}s, 74001-970 , Goi\^{a}nia, GO, Brazil}

\cortext[Bufaical]{Corresponding author}

\begin{abstract}
Structural, electronic and magnetic properties of polycrystalline La$_{2-x}$Ca$_{x}$CoMnO$_{6}$ ($0 \leq x \leq 0.75$) compounds are investigated by x-ray diffraction and magnetometry. All the samples have an orthorhombic structure and show a slight decrease in the unit cell with Ca-doping. Temperature-dependent magnetization measurements reveal a complex magnetic behavior with two ferromagnetic transitions. These transitions are ascribed to Co$^{2+}$--Mn$^{4+}$ and Co$^{3+}$--Mn$^{3+}$ couplings and suggest the presence of additional antiferromagnetic couplings in these disordered compounds. Field-dependent magnetization curves, measured after cooling the samples in a zero external magnetic field, reveal the spontaneous exchange bias (SEB) effect for the Ca-doped samples. The strengthening of the uncompensated magnetic coupling at the interfaces, caused by the increased antiferromagnetic phase, explains the increase of SEB with increasing the Ca-content.
\end{abstract}

\begin{keyword}
Double-perovskite; Cobalt; Manganese; Exchange Bias
\end{keyword}

\end{frontmatter}

\section{INTRODUCTION}

Oxide perovskites are at the frontline of materials research. Its structural flexibility allows the accommodation of several distinct elements on its general formula $AB$O$_3$ [$A$ = alkaline-/rare-earth; $B$ = transition metal (TM)]. This flexibility gives rise to thousands of different compounds already synthesized, many of them displaying properties of fundamental and practical interest, such as superconductivity \cite{Bednorz}, half-metallicity \cite{Serrate} and multiferroicity \cite{Yang}.
The manganites $A$MnO$_3$ are by far the most investigated perovskites due to their metal-insulator transition, accompanied by colossal magnetoresistance \cite{Salamon}. The intense research on La$_{1-x}A_x$MnO$_3$ mixed-valence manganites has led to the formulation of critical physical concepts as double-exchange interactions and Jahn-Teller polarons \cite{Coey}. The cobaltites, $A$CoO$_3$, are also of great significance because the delicate balance between the crystal field splitting and interatomic Coulomb repulsion for Co electrons in octahedral coordination can lead to distinct spin states for this ion in perovskites, say high spin (HS), low spin (LS) or even intermediate spin (IS) configuration \cite{Raveau}. LaCoO$_3$ for instance exhibits temperature-induced spin-state transition \cite{Cheong}, while hole-doped La$_{1-x}$Sr$_x$CoO$_3$ compounds show exchange bias (EB) effect \cite{Tang}, which is an unidirectional anisotropy (UA) set at the magnetic interfaces of heterogeneous systems, manifested as a shift of the magnetization hysteresis loop along field axis when the system is cooled down in an external magnetic field ($H$) \cite{Nogues}.

It is thus not surprising that the combined effect of Co and Mn in the $A_2$CoMnO$_6$ double-perovskites (DP) leads to exciting physical properties. For instance, there is a great interest in the ferromagnetic (FM) insulator La$_2$CoMnO$_6$ compound due to its near room-temperature magnetodielectric effect \cite{Lin,Murthy3}. In contrast, the alkaline earth-doped La$_{1.5}$Sr$_{0.5}$CoMnO$_6$ compound shows a robust spontaneous EB (SEB) effect \cite{Murthy}, a phenomenon in which UA is observed even after cooling the system in zero $H$, being so also called Zero Field Cooled EB effect \cite{Wang,Nayak}. 

The SEB effect seems to be intrinsically related to the presence of a spin glass (SG) phase resulting from the competing magnetic phases in disordered materials \cite{Model,Model2}. For the case of CoMn-based DPs, we previously showed that replacing Sr by Ba or Ca in La$_{1.5}$Sr$_{0.5}$CoMnO$_6$ also leads to SEB. In that case, the Ca-based sample exhibits a small UA while for the Ba-based one, the magnitude of the effect is intermediate between those of Ca- and Sr-based compounds. Such pronounced changes in the SEB effect from one sample to another can be attributed to different proportions of Co$^{2+}$/Co$^{3+}$ and to the spin state of Co$^{3+}$, that affect the uncompensated antiferromagnetic (AFM) Co--Mn coupling \cite{JMMM2017,PRB2019,BJP2020}. Interestingly, single-crystalline La$_{2-x}A_x$CoMnO$_6$ ($A$ = Ca, Sr) samples are not reentrant SG and showed no trace of SEB effect, giving further evidence that an SG-like phase is necessary for the emergence of SEB \cite{PRM2021}

Very recently, a thorough investigation of the La$_{1.5}$(Sr$_{0.5-x}$Ba$_x$)CoMnO$_6$ series has evidenced how the crystal structure affects the SEB effect by indirectly altering the systems' electronic structure \cite{APL2020}. The lattice expansion due to Sr$^{2+}$ to Ba$^{2+}$ partial substitution increases the portion of HS Co$^{3+}$, which by one hand strengthens the Co$^{3+}$--Mn$^{4+}$ interaction, and by the other, it decreases the portion of uncompensated spins ate the magnetic interfaces. The combined effects of enhanced AFM phase and decreased uncompensation in the Co--Mn coupling causes the intermediate region of the series ($x\simeq0.25$) to present the most significant SEB effect reported so far.

Here, we focus our attention on how the immediate changes performed in the electronic configurations of Co and Mn by changing its formal valences impact the SEB of CoMn-based DPs. Our system of interest is La$_{2-x}$Ca$_x$CoMnO$_6$ ($x$ = 0, 0.25, 0.5, 0.75), for which the proximity between Ca$^{2+}$ and La$^{3+}$ ionic radii at XII coordination \cite{Shannon} makes the structural changes caused by the chemical substitution minimal, allowing us to ascribe the evolution of the SEB in this series primarily to changes in Co/Mn formal valences. We observe a systematic increase of the SEB field with increasing the Ca to La substitution, which is discussed in terms of the concomitant increase in the fraction of Co$^{3+}$ present in the system.

\section{EXPERIMENT DETAILS}

Polycrystalline samples of La$_{2-x}$Ca$_{x}$CoMnO$_{6}$ series ($x$ = 0, 0.25, 0.5, 0.75) were synthesized by conventional solid-state reaction method. Stoichiometric amounts of La$_{2}$O$_{3}$, CaO, Co$_{3}$O$_{4}$ and MnO were mixed and heated at $800^{\circ}$C for 12 hours in air atmosphere. Later the samples were re-grinded before the second step of 24 hours at $1200^{\circ}$C. Finally, each sample was ground, pressed into a pellet, and heated at $1300^{\circ}$C for 24 hours. After this procedure, dark-black materials were obtained in 10 mm diameter disks. Attempt to produce the $x$ = 1.0 concentration using this synthesis route was unsuccessful, resulting in impurity phases.

High-resolution x-ray powder diffraction (XRD) data were collected for each sample at room temperature using a PANalytical Empyrean diffractometer, operating with Cu $K_{\alpha}$ radiation. The XRD data was investigated over the angular range $10^{\circ}\leq 2\theta\leq90^{\circ}$, with a 2${\theta}$ step size of 0.013$^{\circ}$. Rietveld refinement was performed with GSAS software, and its graphical interface program \cite{GSAS}.

Magnetization ($M$) as a function of temperature [$M(T)$] and $M$ as a function of $H$ [$M(H)$] measurements were carried out in both zero field cooled (ZFC) and field cooled (FC) modes, using a Quantum Design PPMS-VSM magnetometer. For the ZFC experiments, it was given particular care to eliminate any small trapped field in the magnet, with the samples being demagnetized with an oscillating field at room temperature from one measurement to another, in order to exclude the possibility of the shift in the $M(H)$ curves be related to instrumental artifact.

\section{RESULTS AND DISCUSSION}

\begin{figure}
\begin{center}
\includegraphics[width=0.7 \textwidth]{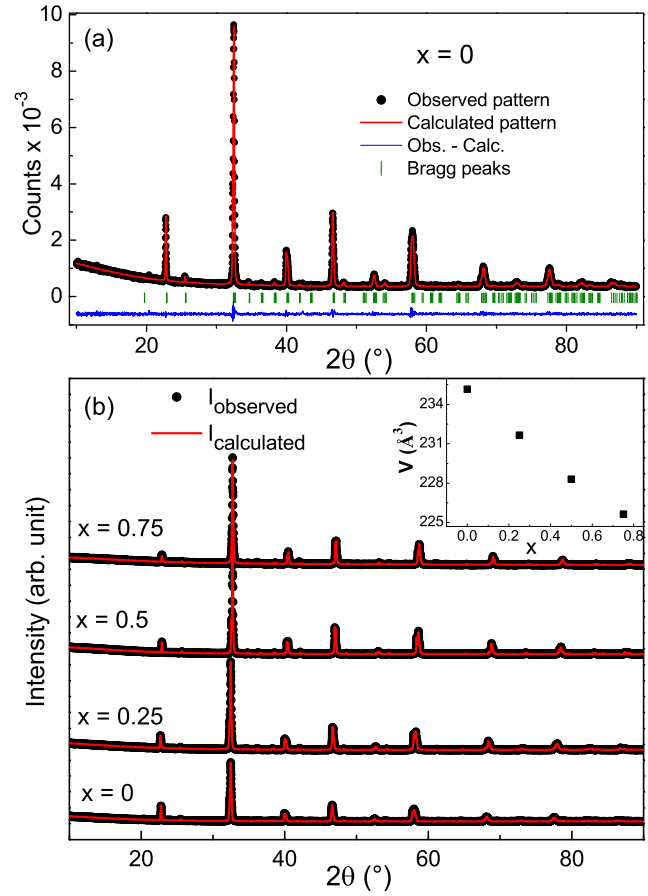}
\end{center}
\caption{(a)Rietveld refinement fitting of La$_2$CoMnO$_{6}$ sample. The solid black circles represent the experimental data, the red solid lines are for the calculated patterns, the blue line shows the difference between the observed and calculated patterns, and the dark green vertical lines represent the Bragg reflections for the $Pnma$ space group. (b) Comparison between the XRD patterns of all La$_{2-x}$Ca$_{x}$CoMnO$_{6}$ samples. Inset shows the unit cell volume as a function of Ca-concentration, $x$.}
\label{Fig_XRD}
\end{figure}

The XRD patterns of La$_{2-x}$Ca$_{x}$CoMnO$_{6}$ samples here investigated (hereafter called by its Ca-concentration, $x$) are shown in Fig. \ref{Fig_XRD}. There is an open debate concerning whether these compounds form in the monoclinic $P2_1/n$ or the orthorhombic $Pnma$ space group. In practice, the main difference between these space groups is that the Co and Mn ions are expected to alternate orderly inside the oxygen octahedra for the monoclinic structure. In contrast, the cations are disordered along the lattice for the orthorhombic one. The controversy occurs because the Co and Mn scattering factors are very similar for XRD performed with Cu-$K_{\alpha}$ radiation. Different groups have reported such compounds as belonging to one of these space groups or even a mixture of both \cite{PRB2019,Sahoo,Xu,Sahoo3,Xu2,Troyanchuck}. The Co- and Mn-based perovskites are known to be very sensitive to the synthesis route \cite{Raveau,Latif,Obradors}, and the La$_{2-x}$Ca$_{x}$CoMnO$_{6}$ seems to be a clear manifestation of this. In our case, the XRD patterns indicate the formation of single-phase perovskite for every investigated sample. 

Fig. \ref{Fig_XRD}(a) shows the Rietveld refinement fitting of the $x$ = 0 compound as representative of the whole series, for which nearly the same overall quality of refinement was achieved. For the adjustment of the calculated patterns, the lattice parameters, the thermal displacements parameters, and the atomic positions were refined, as well as the peak shapes, for which it was used a pseudo-Voigt function, the $U$, $V$ and $W$ terms of the Cagliotti's function being refined for the Gaussian part of the function, and the $X$, $Y$ terms being refined for the Lorentzian one. In order to avoid divergence in the calculations and/or unreliable results, the isotropic temperature factors of Co and Mn were constrained to have the same value, as well as those of oxygen ions. The main results obtained from the refinements for the orthorhombic $Pnma$ space group are displayed in Table \ref{T1}.

As Fig. \ref{Fig_XRD}(b) shows, the diffraction peaks could be well indexed in the $Pnma$ structure for all compounds. The maintenance of the crystallographic symmetry along the series is not surprising since the La$^{3+}$ and Ca$^{2+}$ ionic radii are very close in XII coordination (1.36 \AA and 1.34 \AA, respectively \cite{Shannon}). However, it is important to point out that our laboratory XRD results may have not enough resolution to unambiguously determine the crystallographic structure of this system, the formation in the $P2_1/n$ space group, or even phase segregation, with the presence of some small amount of a secondary phase, can not be excluded.

\begin{table*}
\centering
\caption{Results of Rietveld refinement of crystalline structure for La$_{2-x}$Ca$_{x}$CoMnO$_{6}$ samples in $Pnma$ space group. The unit cell is centered on the Co/Mn ions, so their atomic position is (0,0,0).}
\label{T1}
\begin{tabular}{c|cccc}
\hline
\hline
Sample & 0 & 0.25 & 0.5 & 0.75 \\
\hline
$a$ (\AA) & 5.4832(3) & 5.4580(1) & 5.4335(1) & 5.4153(1) \\
$b$ (\AA) & 7.7685(3) & 7.7246(2) & 7.6869(1) & 7.6587(2) \\
$c$ (\AA) & 5.5206(3) & 5.4950(1) & 5.4661(1) & 5.4405(1) \\
V (\AA$^{3}$) & 235.16(3) & 231.67(1) & 228.31(1) & 225.64(1) \\
\hline
  & \multicolumn{4}{c}{Atomic Positions} \\ 
La/Ca $x$ & 0.5224(2) & 0.5195(3) & 0.5191(3) & 0.5186(5) \\
La/Ca $y$ & 0.25 & 0.25 & 0.25 & 0.25 \\ 
La/Ca $z$ & 0.0038(6) & 0.0027(6) & 0.0032(6) & 0.0013(10) \\
O$_1$ $x$ & -0.0083(22) & -0.0101(24) & -0.0137(24) & -0.0211(34) \\
O$_1$ $y$ & 0.25 & 0.25 & 0.25 & 0.25 \\  
O$_1$ $z$ & -0.0575(24) & -0.0590(20) & -0.0528(21) & -0.0719(28) \\
O$_2$ $x$ & 0.2809(30) & 0.2679(28) & 0.2541(26) & 0.2570(30) \\
O$_2$ $y$ & 0.0371(19) & 0.0343(12) & 0.0331(12) & 0.0264(16) \\
O$_2$ $z$ & 0.2187(13) & 0.2129(17) & 0.2038(15) & 0.2103(21) \\
\hline
  & \multicolumn{4}{c}{Bonds and Angles} \\ 
Co/Mn--O$_1$ (\AA) & 1.968(2) & 1.959(2) & 1.945(2) & 1.956(1) \\
Co/Mn--O$_2$ (\AA) & 1.978(1) & 1.891(1) & 1.792(1) & 1.788(2) \\
Co/Mn--O$_2$ (\AA) & 1.985(1) & 2.041(1) & 2.115(1) & 2.087(1) \\       
$\langle$Co/Mn--O$\rangle$ (\AA) & 1.977(2) & 1.964(2) & 1.951(1) & 1.944(1) \\
Co--O$_1$--Mn ($^{\circ}$) & 161.3(6) & 160.7(6) & 162.4(7) & 156.3(2) \\
Co--O$_2$--Mn ($^{\circ}$) & 158.1(6) & 160.0(5) & 161.0(5) & 164.3(1) \\ 
$\langle$Co--O--Mn$\rangle$ ($^{\circ}$) & 159.7(6) & 160.4(6) & 161.7(6) & 160.3(2) \\
\hline
$R_{wp}$ ($\%$) & 4.7 & 3.9 & 3.8 & 4.1 \\
$\chi^{2}$ & 1.2 & 1.3 & 1.3 & 1.7 \\
\hline
\hline
\end{tabular}
\end{table*}

A careful inspection of the XRD data reveals a systematic shift of the diffraction peaks to the right with increasing the Ca content, related to the lattice shrinkage as can be seen on the inset of Fig. \ref{Fig_XRD}(b). This shift could be ascribed to the fact the Ca$^{2+}$ ionic radius is slightly smaller than that of La$^{3+}$. However, it is also related to the systematic increase in the oxidation states of Co and Mn, for which the ionic radii are expected to decrease as the formal valences increase \cite{Shannon}. Such changes are also manifested in the average Co/Mn--O bond length and directly impact the systems' magnetic properties, as will be discussed next.

\begin{figure}
\begin{center}
\includegraphics[width=0.7 \textwidth]{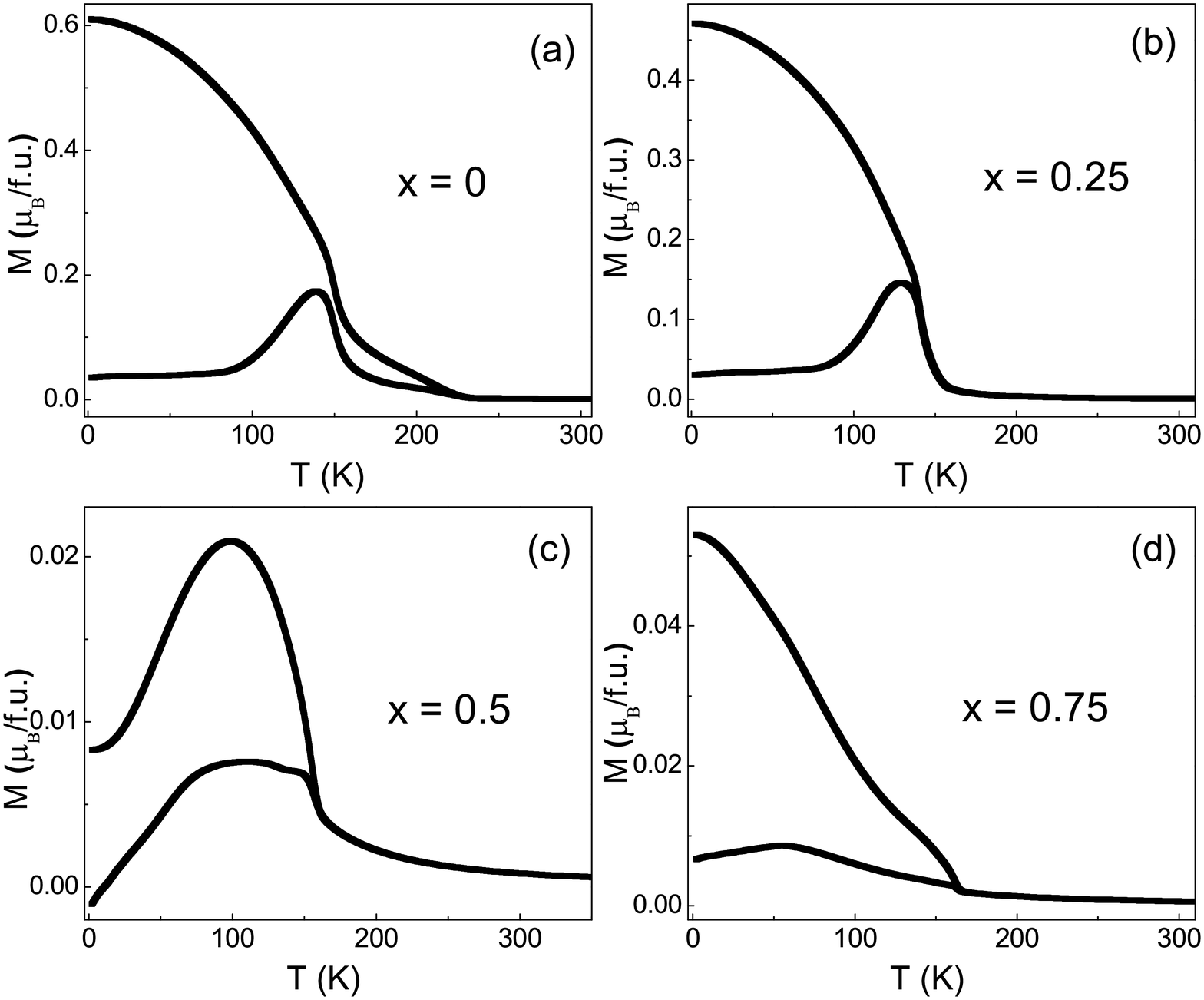}
\end{center}
\caption{ZFC and FC $M(T)$ curves of La$_{2-x}$Ca$_{x}$CoMnO$_{6}$ samples, measured at $H$ = 100 Oe.}
\label{Fig_MxT}
\end{figure}

The ZFC-FC $M(T)$ measurements were performed with $H$ = 100 Oe for every sample. Fig. \ref{Fig_MxT}(a) shows the curves for $x$ = 0, where the presence of two FM-like transitions is clear, in agreement with previous reports for this compound \cite{PRB2019,Dass}. The higher temperature magnetic ordering ($T_{C1}$) is usually ascribed to Co$^{2+}$--O--Mn$^{4+}$ coupling, which is predicted by the Goodenough-Kanamori-Anderson rules to be of FM type, while the lower temperature one ($T_{C2}$) is proposed to come from Co$^{3+}$--O--Mn$^{3+}$ vibronic FM coupling \cite{Dass,GKA}. There are other possible scenarios, as for instance the well known Mn$^{3+}$--O--Mn$^{4+}$ FM coupling \cite{Salamon,Coey} which is possible due to the antisite disorder (ASD) at Co/Mn site, intrinsic of the crystallographic structure here observed for this DP. In any case, the presence of two FM transitions makes evident that mixed-valence states Co$^{2+}$/Co$^{3+}$ and Mn$^{4+}$/Mn$^{3+}$ are present already in our $x$ = 0 sample, as commonly observed for La$_2$CoMnO$_6$ samples produced with similar synthesis routes \cite{PRB2019,Dass,Fournier}.

\begin{figure}
\begin{center}
\includegraphics[width=0.8 \textwidth]{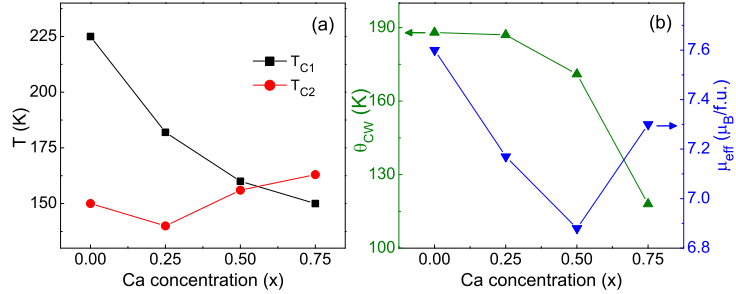}
\end{center}
\caption{(a) $T_{C1}$ and $T_{C2}$ extracted from the first derivatives of the ZFC $M(T)$ curves. (b) $\theta_{CW}$ and $\mu_{eff}$ obtained from the fitting of the PM regions of the $M(T)$ curves with the CW law. Straight lines are guides to the eye.}
\label{Fig_Tc}
\end{figure}

For the Ca-doped samples [Figs. \ref{Fig_MxT}(b)-(d)], besides $T_{C1}$ and $T_{C2}$ it can be noticed a subtle anomaly in the ZFC curves at lower temperatures, in particular for $x$ = 0.5 and 0.75 samples, which is usually related to the presence of AFM couplings like Co$^{2+}$--Co$^{2+}$, Co$^{3+}$--Co$^{3+}$, Mn$^{3+}$--Mn$^{3+}$, Mn$^{4+}$--Mn$^{4+}$ and Co$^{3+}$--Mn$^{4+}$. The AFM exchange interaction may be expected for the $Pnma$ space group where the Co/Mn ions are disorderly distributed along the lattice. The $T_{C1}$ and $T_{C2}$ values, defined here by the local minima in the first derivative of the ZFC curves (not shown), are displayed in Table \ref{T2} and in Fig. \ref{Fig_Tc}(a). The subtle anomalies associated with the AFM couplings are partially masked in the curves due to the stronger FM interactions, hampering the estimate of $T_N$ from our dc measurements, which is thus not displayed here. 

Fig. \ref{Fig_Tc}(a) shows a systematic decrease of $T_{C1}$ with increasing the Ca-content while the changes in $T_{C2}$ are non-monotonic. Although the ion’s electronic configurations primarily influence the magnetic coupling between Co and Mn, other parameters such as the bond lengths and angles may play an important role. The Co--O--Mn coupling in perovskites is generally understood in terms of the $e_g$--$e_g$ orbital hybridization, but the $t_{2g}$--$t_{2g}$ interaction is also expected to play its part in non-cubic  3$d$ perovskites due to the presence of half-filled $t_{2g}$ orbitals in both Co and Mn ions \cite{Goodenough,Blasse}. In this sense, the decrease of $T_{C1}$ can be ascribed mainly to the reduction of the Co$^{2+}$--O--Mn$^{4+}$ phase due to hole doping. 

\begin{table}
\centering
\caption{Main parameters obtained from the $M(T)$ and $M(H)$ measurements.}
\begin{tabular}{c|cccc}
\hline
\hline
Sample & 0 & 0.25 & 0.5 & 0.75 \\
\hline
$T_{C1}$ (K) &  225 & 182 & 160 & 150 \\
$T_{C2}$ (K) & 150 & 140 & 156 & 163 \\
$\theta_{CW}$ (K) & 188 & 187 & 171 & 118 \\
$\mu_{eff}$ ($\mu_B$/f.u.) & 7.6 & 7.2 & 6.9 & 7.3 \\
\hline
$H_{EB}$ (Oe) & - & 70 & 230 & 670 \\
$H_{C}$ (kOe) & 10.2 & 11.3 & 6.8 & 6.3 \\
$M_r$ ($\mu_B$/f.u.) & 2.3 & 2.0 & 0.4 & 0.2 \\
$M_s$ ($\mu_B$/f.u.)  & 4.2 & 4.0 & 2.7 & 1.9 \\
\hline
\hline
\end{tabular}
\label{T2}
\end{table}

In the case of $T_{C2}$, the Ca-doping also strengths the $t_{2g}$--$t_{2g}$ interaction due to the additional hole in the Co$^{3+}$ $t_{2g}$ orbitals. This $\pi$ bonding is expected to be weakened by decreasing the tilts and rotations of the oxygen octahedra, \textit{i.e.}, with increasing the Co--O--Mn bond angle. Table \ref{T2} shows a decrease of $T_{C2}$ followed by an increase with Ca-doping, while the opposite trend is observed for the Co--O--Mn angle. This may be the primary parameter to account for the non-systematic changes in $T_{C2}$. However, other ingredients such as the presence of Co--Co and Mn--Mn interactions induced by the cationic disorder can not be ruled out. Finally, we must stress that the proximity between $T_{C1}$ and $T_{C2}$ may be preventing a precise determination of the latter transition temperature from the first derivative of our dc $M(T)$ curves. Other experiments such as ac magnetic susceptibility would be necessary to accurately determine the magnetic transition temperatures.

It is important to reinforce the sample dependence observed for these CoMn-based perovskites. The widely investigated $x$ = 0 compound is a prime example of that, where depending on the synthesis route, the sample can display one, two, or even three FM transitions. In addition, the AFM couplings are usually present due to the disorder and even possibly a glassy magnetic state is found, resulting from the competition between magnetic phases \cite{Lin,Murthy3,PRB2019,Dass,Fournier}. This sample dependence is also evident for the $x$ = 0.5 concentration, where the magnetic properties here described are similar to those found for samples also produced by solid-state reaction \cite{JMMM2017,PRB2019}. In contrast, the second FM transition is not observed for samples produced by the sol-gel technique, and the sample develops an anomaly at lower temperatures which is ascribed to canted AFM. Additionally, the reentrant glassy behavior can be classified as canonical SG or cluster SG (CG) depending on the synthesis condition \cite{Sahoo,Sahoo2}. This is usual for Mn- and Co-based perovskites \cite{Salamon,Raveau}, the La$_{2-x}$Sr$_x$CoMnO$_6$ system being also very sample-dependent \cite{Murthy,PRB2019}.

From the fitting of the paramagnetic (PM) region of the $M(T)$ curves with the Curie-Weiss (CW) law, we get the CW temperature, $\theta_{CW}$, and the effective magnetic moment, $\mu_{eff}$, for every sample here investigated, as displayed in Table \ref{T2} and Fig. \ref{Fig_Tc}(b). For $x$ = 0, we have $\theta_{CW}$ = 188 K. This largely positive value indicates that FM coupling dominates, but the fact it is somewhat smaller than $T_{C1}$ = 225 K suggests the presence of AFM phases already for the parent compound, as expected for this disordered system.

We can infer the Co/Mn electronic configurations for the $x$ = 0 compound by comparing its $\mu_{eff}$ = 7.6 $\mu_B$/f.u. experimentally observed with the value obtained from the usual equation for systems with two or more different magnetic ions \cite{Shin,PRB2020}
\begin{equation}
\mu= \sqrt{{\mu_1}^2 + {\mu_2}^2 + {\mu_3}^2 +...} \label{Eq1}
\end{equation}
As aforementioned, one may expect mixed valences for the TM ions already for the $x$ = 0 parent compound. Assuming the same proportion of Co$^{2+}$/Mn$^{4+}$ and Co$^{3+}$/Mn$^{3+}$, considering that Co$^{3+}$ is in HS configuration and using the spin only (SO) approximation for these 3$d$ ion moments ($\mu_{Co^{2+}}$ = $\mu_{Mn^{4+}}$ = 3.9 $\mu_B$; $\mu_{HS Co^{3+}}$ = $\mu_{Mn^{3+}}$ = 4.9 $\mu_B$ \cite{Ashcroft}) we get $\mu$ = 6.2 $\mu_B$/f.u., far below the experimental value. One of the main reasons for such large discrepancy may be the fact that some orbital contribution for the magnetic moments should be expected, in special for Co ions \cite{Raveau,PRB2019}. Alternatively, considering the same proportion of mixed valence ions but now using the standard magnetic moments usually found for these 3$d$ TM ions ($\mu_{Co^{2+}}$ = 4.8 $\mu_B$; $\mu_{HS Co^{3+}}$ = 5.4 $\mu_B$; $\mu_{Mn^{3+}}$ = 5 $\mu_B$; $\mu_{Mn^{4+}}$ = 4 $\mu_B$ \cite{Ashcroft}) we get $\mu$ = 6.8 $\mu_B$/f.u., still smaller but now much closer to the experiment. This calculation is undoubtedly just a rough estimate because the precise orbital moment of each ion is unknown, and the exact Co$^{2+}$/Co$^{3+}$ and Mn$^{3+}$/Mn$^{4+}$ ratios may be somewhat different from that assumed in La$_{2}$(Co$^{2+}_{0.5}$Co$^{3+}_{0.5}$)(Mn$^{3+}_{0.5}$Mn$^{4+}_{0.5}$)O$_6$. Besides, one may also have some fraction of Co$^{3+}$ in LS configuration, as already found for similar DPs \cite{PRB2019,APL2020,Vashook}. A detailed investigation employing X-ray absorption spectroscopy, X-ray photoelectron spectroscopy, X-ray magnetic circular dichroism and/or neutron diffraction is necessary to precisely determine each ion's formal valence and magnetic moment.

Fig. \ref{Fig_Tc}(b) and Table \ref{T2} show that for the  $x$ = 0 - 0.5 interval the $\mu_{eff}$ decreases as the Ca-content increases. Since Mn$^{4+}$ moment is smaller than that of Mn$^{3+}$ whereas the one of HS Co$^{3+}$ is larger than that of HS Co$^{2+}$, this decrease in $\mu_{eff}$ strongly suggests that up to $x$ = 0.5 the hole doping acts mainly to increase the Mn formal valence. Otherwise, the effective moment would increase. Assuming thus for $x$ = 0.25 and 0.5 the respective general formulas La$_{1.75}$Ca$_{0.25}$(Co$^{2+}_{0.5}$Co$^{3+}_{0.5}$)(Mn$^{3+}_{0.25}$Mn$^{4+}_{0.75}$)O$_6$ and La$_{1.5}$Ca$_{0.5}$(Co$^{2+}_{0.5}$Co$^{3+}_{0.5}$)Mn$^{4+}$O$_6$ in Eq. \ref{Eq1} we get the theoretical moments $\mu$ = 6.7 $\mu_B$/f.u. and $\mu$ = 6.5 $\mu_B$/f.u., respectively, thus capturing the decrease of $\mu_{eff}$ with increasing $x$ that is observed experimentally.

Interestingly, for $x$ = 0.75 it is observed the increase of $\mu_{eff}$ with respect to the value found for $x$ = 0.5. This rise is because, for $x$ = 0.5, all the Mn ions are already at the tetravalent state, and Mn$^{5+}$ is very unlikely. Thus the further increment of Ca$^{2+}$ at La$^{3+}$ site from $x$ = 0.5 to 0.75 now acts to remove electrons from Co, increasing its average oxidation state. Assuming La$_{1.25}$Ca$_{0.75}$(Co$^{2+}_{0.25}$Co$^{3+}_{0.75}$)Mn$^{4+}$O$_6$ for this sample in Eq. \ref{Eq1} we get $\mu$ = 6.6 $\mu_B$/f.u., which represents an increase in comparison to the theoretical value obtained for $x$ = 0.5 and again being in accordance with the experimental results. Again, the discrepancies between the theoretical and experimental values indicate an oversimplification of the problem. The presence of two magnetic transitions for the $x$ = 0.5 and 0.75 samples indicate that one may have, in practice, some small fraction of Mn$^{3+}$ even for these compounds. Such simplified calculations do not intend to give the precise estimate of the proportion between the TM ions but only to give us some insight about the trends induced by Ca-doping. Although the calculated values are somewhat smaller than those obtained experimentally, the trend observed with changing the Ca-concentration could be captured precisely.

\begin{figure*}
\begin{center}
\includegraphics[width=1.3 \textwidth]{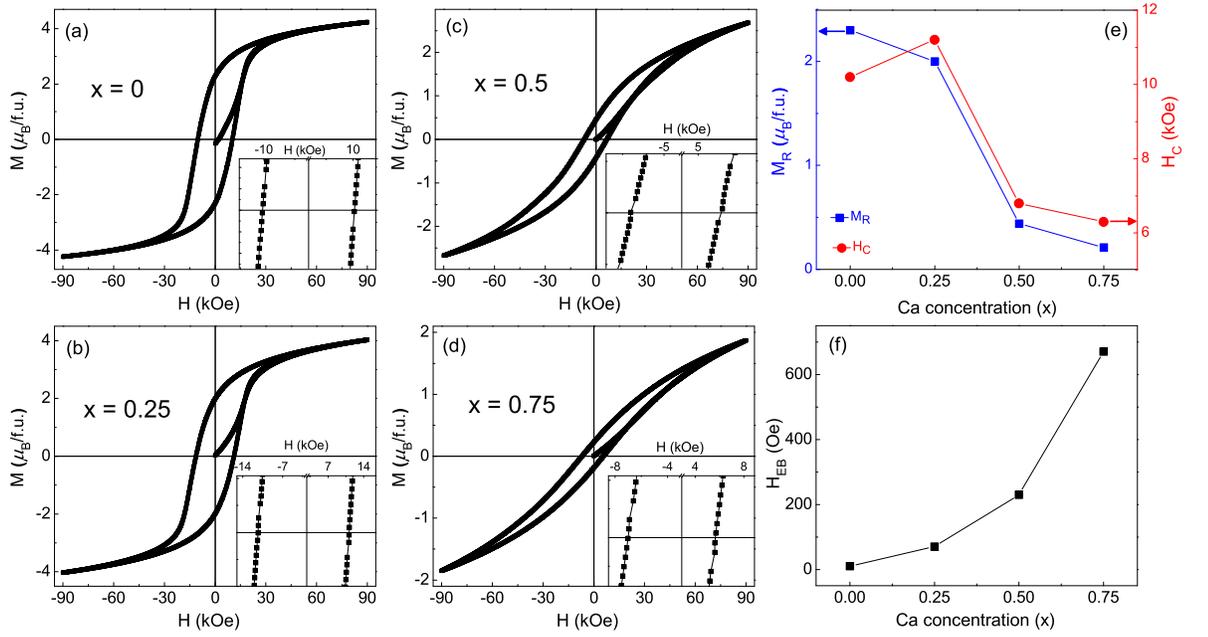}
\end{center}
\caption{(a)-(d) $M(H)$ loops for La$_{2-x}$Ca$_{x}$CoMnO$_{6}$ measured at 5 K after ZFC each sample. Insets show magnified views of the regions near $M$ = 0, highlighting the horizontal shifts in the Ca-doped samples. (e) $M_R$ and $H_C$ as a function of Ca-concentration, $x$. (f) $H_{EB}$ as a function of $x$. Straight lines are guides to the eye.}
\label{Fig_MxH}
\end{figure*}

Fig. \ref{Fig_MxH} shows the $M(H)$ curves measured up to $H$ = 90 kOe after ZFC each sample down to 5 K. As can be noticed, all loops are closed, nearly vertically symmetric, and show some hysteresis. Notably, the loops gradually lose their squareness with increasing the Ca-concentration and develop a linear field-dependence at high fields, suggesting the increase in the proportion of AFM phase with increasing $x$. This change in the shape of the hysteresis curves is also manifested in the systematic decrease of the remanent magnetization ($M_R$), as shown in Fig. \ref{Fig_MxH}(e). In addition, a similar trend is observed for the coercive field $H_{C}=(H_{C+}-H_{C-})/2$, where $H_{C+}$ and $H_{C-}$ are respectively the positive and negative coercive fields, which will have a direct impact on the SEB effect.

The shift in the $M(H)$ curves, characteristic of the EB effect, are not large enough for the La$_{2-x}$Ca$_{x}$CoMnO$_{6}$ samples in a way to be apparent in the complete hysteresis loops shown in the main Figs. \ref{Fig_MxH}(a)-(d), but are evidenced at the magnified views of the regions close to $M$ = 0 displayed in the insets. As can be seen, there is no shift for the $x$ = 0 parent compound, but for the Ca-doped samples, the asymmetry in the curves is clear, with the EB field, $H_{EB}=|H_{C+}+H_{C-}|/2$, systematically increasing with $x$, as shown in Fig. \ref{Fig_MxH}(f) and Table \ref{T2}.

The increase of $H_{EB}$ with increasing the $A^{2+}$ doping at La$^{3+}$ site observed here for La$_{2-x}$Ca$_{x}$CoMnO$_{6}$ is different to the trends reported for La$_{2-x}$Sr$_{x}$CoMnO$_{6}$ and La$_{1.5}$(Sr$_{0.5-x}$Ba$_{x}$)CoMnO$_{6}$ systems, for which the SEB field is larger at intermediate $x$ values. For the first series, this trend is directly related to the ASD between the TM ions, which destroys the long-range magnetic order and increases the AFM interactions. The canted AFM phase observed at low temperature is larger for $x$ = 0.5. Correspondingly, the SEB is also maxima at this concentration \cite{Murthy2}. For La$_{1.5}$(Sr$_{0.5-x}$Ba$_{x}$)CoMnO$_{6}$ the lattice expansion due to Ba to Sr substitution leads to the increase in the concentration of HS Co$^{3+}$ ions which in turn leads to two competing effects, the increase of the AFM phase and the reduction of the uncompensation in this AFM coupling. This combined effect results in a maximum in the SEB field for the $x$ = 0.25 intermediate concentration of Ba/Sr \cite{APL2020}.

A previous investigation of the SEB effect in La$_{2-x}$Ca$_{x}$CoMnO$_{6}$ polycrystals prepared by the sol-gel method \cite{Sahoo3} showed $H_{EB}$ values remarkably different from ours, further demonstrating the severe sample-dependence in this system. Moreover, in contrast to our work, increasing $H_{EB}$ for 0 $\leq x \leq$ 0.5 samples followed by a decrease for $x$ $>$ 0.5 was observed, although the $x$ = 0.75 concentration was not investigated. The authors interpreted such changes in terms of the Co/Mn ASD estimated from the $M_s$ values. In our case, however, in principle we should expect a systematic decrease of ASD in the 0 $\leq x \leq$ 0.5 region due to the increased portion of Mn$^{4+}$ ions, whose radius and valence are more distant to those of Co$^{2+}$/Co$^{3+}$ than Mn$^{3+}$. These parameters are known to play an important role in the ASD of DPs \cite{Serrate,Sami}. 

According to the arguments proposed in Ref. \cite{Sahoo3}, the $H_{EB}$ of our series should decrease for 0 $\leq x \leq$ 0.5, in contrast to the results. Thus, for our La$_{2-x}$Ca$_{x}$CoMnO$_{6}$ samples, we ascribe the increase of $H_{EB}$ directly to the changes in the AFM/FM interfaces. The systematic decrease of $T_{C1}$ and $\theta_{CW}$ with increasing $x$ observed in the $M(T)$ curves, and the decrease of $M_r$ and $M_s$ observed in the $M(H)$ loops, suggest the increase in the portion of the AFM phase with increasing the Ca-content. This strengthens the AFM/FM coupling, enhancing the SEB effect. The loss of squareness of the $M(H)$ loops with increasing $x$ and the changes in the shape of the curves are also suggestive of the presence of a glassy magnetic phase for the Ca-doped samples \cite{Mydosh}, as was already observed for the $x$ = 0.5 compound \cite{JMMM2017,Sahoo,Sahoo2}, and is also expected for the other concentrations of this disordered system where the magnetic phases compete. The SG-like phase is ubiquitous on the SEB effect observed in DP compounds \cite{Model,Model2}.

Although non-negligible, the SEB here described for the Ca-doped samples is much smaller than those reported for La$_{2-x}$A$_x$CoMnO$_6$ compounds with $A$ = Sr and Ba \cite{Murthy,PRB2019,APL2020}, for which the presence of LS Co$^{3+}$ is believed to uncompensate the Co--Mn AFM coupling and thus enhance the UA responsible for the EB effect. The $M(T)$ data for our Ca-based samples suggest that Co$^{3+}$ is in HS configuration, which possibly leads to a nearly compensated AFM coupling and consequently decreases the SEB effect.

\section{CONCLUSIONS}

In summary, the La$_{2-x}$Ca$_{x}$CoMnO$_{6}$ ($0 \leq x \leq 0.75$) samples here investigated form as single-phase perovskites, the XRD patterns could be successfully refined in the orthorhombic $Pnma$ space group. The magnetic measurements indicate a complex magnetic behavior for all samples, where the presence of Co$^{2+}$/Co$^{3+}$ and Mn$^{3+}$/Mn$^{4+}$ mixed-valence states lead to competing FM and AFM interactions, which may result in glassy magnetic behavior and EB effect. The increased portion of the AFM phase with increasing $x$, as suggested by the $M(T)$ data, can explain the rise of the SEB effect with increasing the Ca-doping as observed in the $M(H)$ curves.

\section{ACKNOWLEDGMENTS}
This work was supported by Conselho Nacional de Desenvolvimento Cient\'{i}fico e Tecnol\'{o}gico (CNPq) [No. 425936/2016-3], Coordena\c{c}\~{a}o de Aperfei\c{c}oamento de Pessoal de N\'{i}vel Superior (CAPES), Funda\c{c}\~{a}o de Amparo \`{a} Pesquisa do Estado de Goi\'{a}s (FAPEG), and Funda\c{c}\~{a}o Carlos Chagas Filho de Amparo \`{a} Pesquisa do Estado do Rio de Janeiro (FAPERJ) [Nos. E-26/202.798/2019, E-26/211.291/2021].


\begin{thebibliography}{99}

\bibitem{Bednorz} J. G. Bednorz and K. A. M\"{u}ller, Z. Phys. B Cond. Mat. 64 (1986) 189–193.

\bibitem{Serrate} D. Serrate, J. M. De Teresa and M. R. Ibarra, J. Phys.: Condens. Matter 19 (2007) 023201.

\bibitem{Yang} H. Liu and X. Yang, Ferroelectrics 507:1, (2017) 69-85.

\bibitem{Salamon} M. B. Salamon and M. Jaime, Rev. Mod. Phys. 73 (2001) 583.

\bibitem{Coey} J. M. D. Coey, M. Viret and S. von Moln\'{a}r, Adv. Phys. 48:2 (1999) 167-293.

\bibitem{Raveau} B. Raveau and Md. Motin Seikh, \textit{Cobalt Oxides: From Crystal Chemistry to Physics} (Wiley-VCH, Weinheim, 2012).

\bibitem{Cheong} P. G. Radaelli and S.-W. Cheong, Phys. Rev. B 66 (2002) 094408.

\bibitem{Tang} Y. Tang, Y. Sun, and Z. Cheng, Phys. Rev. B 73 (2006) 174419.

\bibitem{Nogues} J. Nogu\'{e}s and I. K. Schuller, J. Magn. Magn. Mater. 192 (1999) 203.

\bibitem{Lin} Y. Q. Lin and X. M. Chen, J. Am. Ceram. Soc., 94 [3] (2011) 782–787.

\bibitem{Murthy3} J. Krishna Murthy, K. Devi Chandrasekhar, S. Murugavel, and A. Venimadhav, J. Mater. Chem C 3 (2015) 836.

\bibitem{Murthy} J. Krishna Murthy and A. Venimadhav, Appl. Phys. Lett. 103 (2013) 25410.

\bibitem{Wang} B. M. Wang, Y. Liu, P. Ren, B. Xia, K. B. Ruan, J. B. Yi, J. Ding, X. G. Li, and L. Wang, Phys. Rev. Lett. 106 (2011) 0077203 .

\bibitem{Nayak} A. K. Nayak, M. Nicklas, S. Chadov, C. Shekhar, Y. Skourski, J. Winterlik, and C. Felser, Phys. Rev. Lett. 110 (2013) 127204.

\bibitem{Model} L. T. Coutrim, E. M. Bittar, F. Garcia, and L. Bufai\c{c}al, Phys. Rev. B 98 (2018) 064426.

\bibitem{Model2} L. Bufai\c{c}al, L. T. Coutrim, E. M. Bittar, and F. Garcia,  J. Magn. Magn. Mater. 512 (2020) 167048.

\bibitem{JMMM2017} L. Bufai\c{c}al, R. Finkler, L. T. Coutrim, P. G. Pagliuso, C. Grossi, F. Stavale, E. Baggio-Saitovitch, and E.M. Bittar, J. Magn. Magn. Mater. 433 (2017) 271–277.

\bibitem{PRB2019} L. T. Coutrim, D. Rigitano, C. Macchiutti, T. J. A. Mori, R. Lora-Serrano, E. Granado, E. Sadrollahi, F. J. Litterst, M. B. Fontes, E. Baggio-Saitovitch, E. M. Bittar, and L. Bufai\c{c}al, Phys. Rev. B 100 (2019) 054428.

\bibitem{BJP2020} M. Boldrin, L. T. Coutrim, and L. Bufai\c{c}al, Braz. J. Phys. 50 (2020) 711–715.

\bibitem{PRM2021} C. Macchiutti, J. R. Jesus, F. B. Carneiro, L. Bufai\c{c}al , M. Ciomaga Hatnean, G. Balakrishnan, and E. M. Bittar, Phys. Rev. Mater. 5 (2021) 094402.

\bibitem{APL2020} M. Boldrin, A. G. Silva, L. T. Coutrim, J. R. Jesus, C. Macchiutti, E. M. Bittar, and L. Bufai\c{c}al, Appl. Phys. Lett. 117 (2020) 212402.

\bibitem{Shannon} R. D. Shannon, Acta Crystallographica A32 (1976) 751.

\bibitem{GSAS} A. C. Larson and R. B. Von Dreele, Los Alamos National Laboratory Report No. LAUR 86-748, 2000; B. H. Toby, J. Appl. Crystallogr. \textbf{34}, 210 (2001).

\bibitem{Sahoo} R. C. Sahoo, D.Paladhi, and T. K. Nath, J. Magn. Magn. Mater. 436 (2017) 77-84.

\bibitem{Xu} Q. Li, L. Xing, and M. Xu, Phys. Status Solidi B 254 (2017) 9, 1600757.

\bibitem{Sahoo3} R. C. Sahoo, S. Das, D. Daw, R. Singh, A. Das, and T. K. Nath, J. Phys.: Condens. Matter 33 (2021) 215804.

\bibitem{Xu2} Q. Li, N. Li, J. Hu, Q. Han, Q. Ma, L. Ge, B. Xiao, and M. Xu, J. Appl. Phys. 116 (2014) 033905.

\bibitem{Troyanchuck} I. O. Troyanchuk, A. P. Sazonov, H. Szymczak, D. M. T\"{o}bbens, and H. Gamari-Seale, J. Exp. Theor. Phys. 99 (2004) 363–369.

\bibitem{Latif} I. A. Abdel-Latif, J. Phys. 1 (2012) 3.

\bibitem{Obradors} B. Mart\'{i}nez, Ll. Balcells, J. Fontcuberta, X. Obradors, C. H. Cohenca, and R. F. Jardim, J. Appl. Phys. 83 (1998) 7058.

\bibitem{Vashook} V. Vashook, D. Franke, J. Zosel, L. Vasylechko, M. Schmidt, and U. Guth, J. Alloys Compd. 487 (2009) 577–584.

\bibitem{Dass} R. I. Dass and J. B. Goodenough, Phys. Rev. B 67 (2003) 014401.

\bibitem{GKA} J. B. Goodenough, \textit{Magnetism and Chemical Bond} (Interscience, New York, 1963).

\bibitem{Fournier} K. D. Truong, J. Laverdi\`{e}re, M. P. Singh, S. Jandl, and P. Fournier, Phys. Rev. B \textbf{76}, 132413 (2007).

\bibitem{Goodenough} J. B. Goodenough, A. Wold R. J. Arnott, and N. Menyuk, Phys. Rev. {124}, 2 (1961).

\bibitem{Blasse} G. Blasse, J. Phys. Chem. Solids {26}, 1969-1971 (1965).

\bibitem{Sahoo2} R. C. Sahoo, S. K. Giri, D. Paladhi, A. Das, and T. K. Nath, J. Appl. Phys. 120 (2016) 033906.

\bibitem{Shin} T. Shin-ike, T. Sakai, G. Adachi, and J. Shiokawa, Mat. Res. Bull. 12 (1977) 831-836.

\bibitem{PRB2020} L. Bufai\c{c}al, E. Sadrollahi, F. J. Litterst, D. Rigitano, E. Granado, L. T. Coutrim, E. B. Ara\'{u}jo, M. B. Fontes, E. Baggio-Saitovitch, and E. M. Bittar, Phys. Rev. B 102 (2020) 024436.

\bibitem{Ashcroft} N. W. Ashcroft and N. D. Mermin, \textit{Solid State Physics}, Cengage Learning (1976).

\bibitem{Murthy2} J. Krishna Murthy, K. D. Chandrasekhar, H. C. Wu, H. D. Yang, J. Y. Lin, and A. Venimadhav, J. Phys.: Condens. Matter 28 (2016) 086003.

\bibitem{Sami} S. Vasala and M. Karppinen, Prog. Solid State Chem. 43 (2015) 1.

\bibitem{Mydosh} J. A. Mydosh, \textit{Spin Glasses: An Experimental Introduction} (Taylor \& Francis, London, 1993).

\end{thebibliography}
\end{document}